\documentclass[aip,apl,amsmath,amssymb,reprint,superscriptaddress]{revtex4-2}

\usepackage{graphicx}
\usepackage{siunitx}
\usepackage[colorlinks=true,allcolors=blue]{hyperref}

\graphicspath{{figures/}}

\begin{document}

\title{Electrothermal behavior of superconducting nanowires buried in a
       commercial CMOS process}

\author{Mohamed Gharib}
\email[Authors to whom correspondence should be addressed: ]{mghari5@uic.edu, amiceli@anl.gov and vaisband@uic.edu}
\affiliation{University of Illinois Chicago, Chicago, Illinois 60607, USA}
\affiliation{Argonne National Laboratory, Lemont, Illinois 60439, USA}

\author{Leonid Popryho}
\affiliation{University of Illinois Chicago, Chicago, Illinois 60607, USA}
\affiliation{Argonne National Laboratory, Lemont, Illinois 60439, USA}

\author{Salma Abdelzaher}
\affiliation{University of Illinois Chicago, Chicago, Illinois 60607, USA}
\affiliation{Argonne National Laboratory, Lemont, Illinois 60439, USA}

\author{Tejas Guruswamy}
\affiliation{Argonne National Laboratory, Lemont, Illinois 60439, USA}

\author{Tomas Polakovic}
\affiliation{Argonne National Laboratory, Lemont, Illinois 60439, USA}

\author{Umeshkumar Patel}
\affiliation{Argonne National Laboratory, Lemont, Illinois 60439, USA}

\author{Orlando Quaranta} 
\affiliation{Argonne National Laboratory, Lemont, Illinois 60439, USA}
\affiliation{University of Chicago, Chicago, Illinois 60637, USA}

\author{Thomas Cecil}
\affiliation{Argonne National Laboratory, Lemont, Illinois 60439, USA}

\author{Clarence Chang}
\affiliation{Argonne National Laboratory, Lemont, Illinois 60439, USA}
\affiliation{University of Chicago, Chicago, Illinois 60637, USA}

\author{Yu-Sheng Chen}
\affiliation{University of Chicago, Chicago, Illinois 60637, USA}

\author{Thomas H. Swift}
\affiliation{Quantum Motion, 9 Sterling Way, London, N7 9HJ, United Kingdom}

\author{Grayson M. Noah}
\affiliation{Quantum Motion, 9 Sterling Way, London, N7 9HJ, United Kingdom}

\author{Alberto Gomez-Saiz}
\affiliation{Quantum Motion, 9 Sterling Way, London, N7 9HJ, United Kingdom}
\affiliation{Department of Electrical and Electronic Engineering, Imperial College London, London SW7 2AZ, United Kingdom}

\author{John J. L. Morton}
\affiliation{Quantum Motion, 9 Sterling Way, London, N7 9HJ, United Kingdom}

\author{M. Fernando Gonzalez-Zalba}
\affiliation{Quantum Motion, 9 Sterling Way, London, N7 9HJ, United Kingdom}
\affiliation{CIC nanoGUNE Consolider, Tolosa Hiribidea 76, E-20018 Donostia-San Sebastian, Spain}
\affiliation{IKERBASQUE, Basque Foundation for Science, E-48011 Bilbao, Spain}

\author{Antonino Miceli}
\email[Authors to whom correspondence should be addressed: ]{mghari5@uic.edu, amiceli@anl.gov and vaisband@uic.edu}
\affiliation{Argonne National Laboratory, Lemont, Illinois 60439, USA}
\affiliation{University of Chicago, Chicago, Illinois 60637, USA}

\author{Inna Partin-Vaisband}
\email[Authors to whom correspondence should be addressed: ]{mghari5@uic.edu, amiceli@anl.gov and vaisband@uic.edu}
\affiliation{University of Illinois Chicago, Chicago, Illinois 60607, USA}

\date{\today}

\begin{abstract}

Superconducting films native to a commercial CMOS process enable
co-integration of superconducting devices---high-kinetic-inductance
elements and nanowire detectors---with cryogenic CMOS on one monolithic die.
Such films are buried in the front-end-of-line, beneath the
back-end-of-line metal stack, an environment that might quench a
self-heated hot spot. Current--voltage characteristics are reported
for titanium nitride (TiN) nanowires in the polycrystalline-silicon
resistor layer of a \SI{22}{\nano\metre} FD-SOI process, and measured
retrapping currents are compared with thermal models for longitudinal
conduction to the contacts and interfacial phonon cooling into the
stack. The wires are strongly hysteretic, with switching-to-retrapping
current ratios $I_{sw}/I_r$ of 7.4--9.8, sustaining a normal hot spot.
For two wires of length $\ell=\SI{50}{\micro\metre}$ and widths
$W=\SI{0.50}{\micro\metre}$ and \SI{1.00}{\micro\metre},
over bath temperatures $T_b=\SI{0.15}{\kelvin}$--\SI{1.19}{\kelvin},
$I_r$ follows the interfacial form $(T_{hs}^n-T_b^n)^{1/2}$
with $n=4$ ($T_{hs}$, the effective hot-spot temperature), and is
strongly inconsistent with the $(T_{hs}-T_b)^{1/2}$
constant-$\kappa$ longitudinal-conduction limit; free-exponent fits
give $n\approx4.8$--$5.0$. An effective cooling coefficient
$\Sigma_K=\SI{7.9(3)}{\watt\per\metre\squared\per\kelvin\tothe{4}}$
describes the data: $31(21)\times$ below the acoustic- (diffuse-)
mismatch value for a single clean interface, and $\approx12\times$
below the lowest previously reported comparable nanowire coefficient.
The model-based film thermal conductivity is
$\kappa\approx\SI{20}{\milli\watt\per\metre\per\kelvin}$.
Across lengths $\ell=1$--\SI{50}{\micro\metre},
$I_r$ matches neither $\ell^{-1}$ of pure longitudinal conduction
nor $\ell^{0}$ of pure interfacial cooling; a two-channel model
reproduces $I_r(\ell)$ with crossover length
$\ell^*\approx\SI{3.5}{\micro\metre}$ predicted, to an $O(1)$
prefactor, from $\Sigma_K$ and $\kappa$.
\end{abstract}

\maketitle

Commercial complementary metal oxide semiconductor (CMOS) processes routinely use transition-metal
nitrides and silicides as diffusion barriers, liner metals and passive
components.\cite{wittmer1980,lavoie2003,TES-APS-ICs} Several become superconducting at cryogenic
temperatures\cite{vissers2010,chiu2021, Groll_NICK, CHIU2024348}---a property incidental to processes designed for room-temperature
operation, in which the films are chosen for work function, adhesion and sheet
resistance. Recently Swift \textit{et al.} used one such film as a circuit element.\cite{swift2025} They exploited the high kinetic inductance ($\approx\SI{1}{\nano\henry}$ per square) of the titanium nitride (TiN) sub-film inside a polycrystalline-silicon resistor (poly-resistor) of a GlobalFoundries \SI{22}{\nano\metre} fully depleted silicon-on-insulator (FD-SOI) process. The resulting ``superinductor" is four orders of magnitude more compact in area
than the spiral inductors conventionally available on such a process. It is used to read
out a silicon metal–oxide–semiconductor (MOS) quantum dot in an integrated radio-frequency single-electron transistor, establishing the film as a usable circuit material for
superconductor-enhanced CMOS circuits.

The significance extends beyond passive inductors.
Strongly disordered superconductors of exactly this class (e.g., TiN, NbN, NbTiN, WSi) are
the workhorse materials of superconducting nanowire single-photon detectors\cite{Goltsman, Lita:08, natarajan2012} and of
nanowire logic and memory.\cite{orlando-ntron,Berggren-ntron,htron} They are chosen for the
same properties that make them good superinductors: 
high normal-state resistivity, short coherence length and low
electron diffusivity. A CMOS process that natively contains such a
film is therefore a candidate platform for monolithically integrating nanowire
detectors and superconducting logic and memory circuits with their cryogenic CMOS
readout electronics.

\begin{figure}[tb]
  \includegraphics[width=\columnwidth]{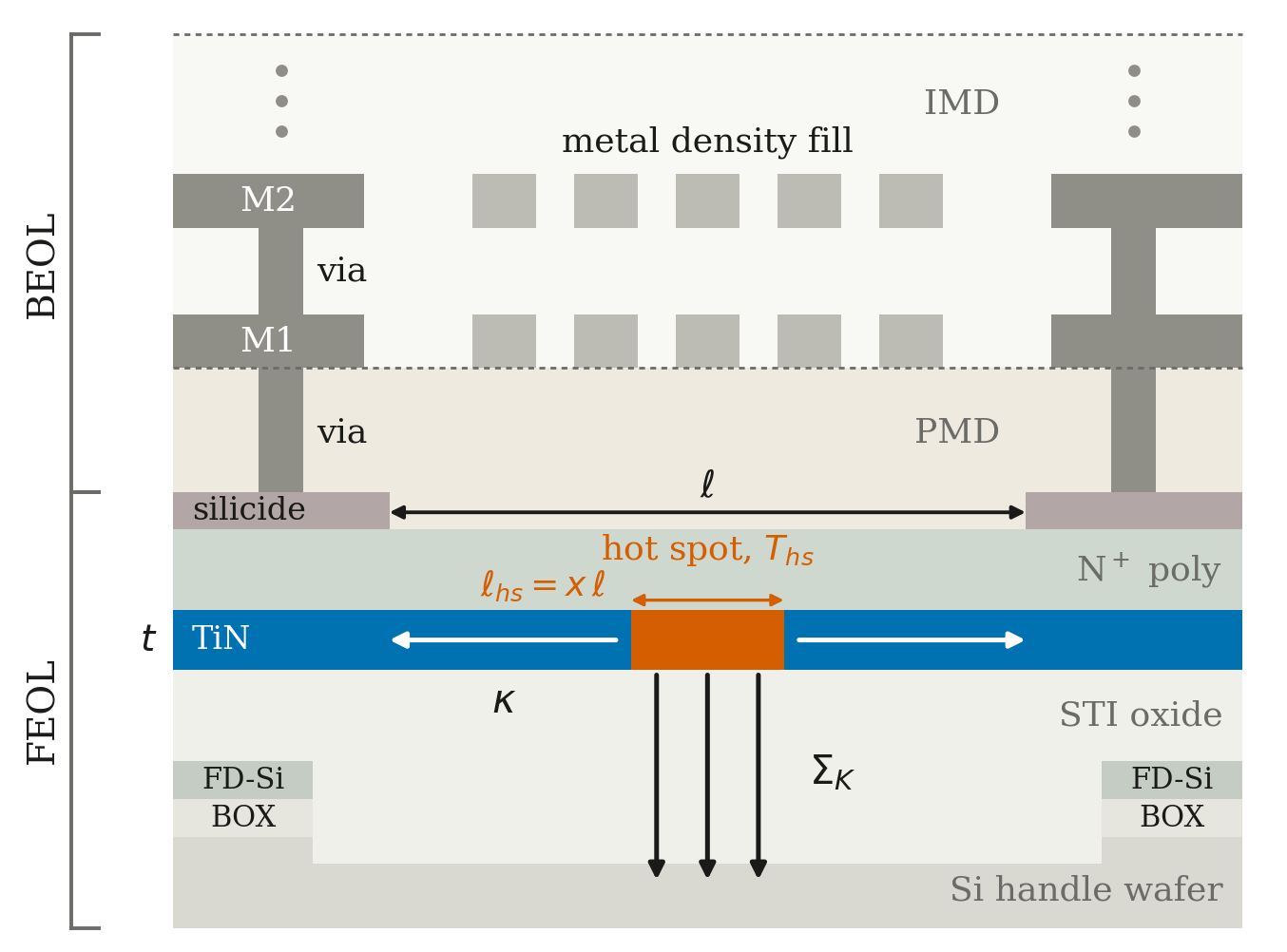}
  \caption{Cross-section along the wire, not to scale. The TiN film lies on the
    shallow-trench-isolation (STI) oxide, is capped by doped polysilicon with silicide
    over the two contact ends, and is buried under the pre-metal dielectric (PMD), the
    BEOL metal levels (M1, M2), and inter-metal dielectric (IMD); isolated M1/M2 blocks
    are floating density fill, and dots mark where the stack continues. A hot spot at
    $T_{hs}$ loses heat along the film to the contacts (set by $\kappa$) and vertically
    into the surrounding layers (set by $\Sigma_K$). $\ell$ is the length between the
    silicide pads; $W$ runs into the page.}
  \label{fig:device}
\end{figure}

The environment of the film matters for that prospect. The front end of line (FEOL) contains the transistor-level structures at the wafer surface---active silicon, isolation, gate stack and silicide---while the back end of line (BEOL) is the multilevel metal-and-dielectric interconnect built on top of it. 
The TiN layer studied here is a few-nanometer-thick film
within the body of a poly-resistor.\cite{DickJames} It lies on shallow-trench-isolation oxide and is capped
by doped polysilicon, silicided
at the  contact ends. Above it the pre-metal dielectric separates the
FEOL from the first metal level of the BEOL stack.\cite{swift2025} The TiN layer is buried in the FEOL beneath the BEOL stack,
encapsulated in amorphous dielectric with oxide below and metal interconnects and density fill above (Fig.~\ref{fig:device}).

The buried environment makes the electrothermal behavior of the TiN film uncertain. In
superconducting nanowire detectors, photon absorption can nucleate a transient resistive region
whose subsequent evolution (e.g., collapse, growth, or latching) is governed by electrothermal feedback;\cite{natarajan2012,kerman2009,annunziata2010}
related feedback is also exploited in nanowire logic and memory.\cite{orlando-ntron,Berggren-ntron,htron}
The retrapping current $I_r$---required to sustain an existing normal (resistive) domain---lies
below the switching current $I_{sw}$ at which a cold superconducting wire becomes resistive.
Whereas $I_{sw}$ is governed by superconducting-instability mechanisms such as depairing, phase
slips, or vortex entry, $I_r$ is determined by heat removal. 
Whether a film buried close to the BEOL metal stack can sustain a hot spot is therefore
not obvious from its geometry alone and requires experimental characterization together
with electrothermal modeling.

Heat can leave the hot spot by two routes, which are considered separately.
A retrapping current $I_r^{\mathrm{cond}}$ is defined for longitudinal
conduction along the film to the contacts, while a distinct retrapping current
$I_r^{\mathrm{int}}$ is defined for interfacial cooling from the film into the
surrounding stack. Unadorned $I_r$ denotes the measured retrapping current.
Because the resistive hot spot is not isothermal, its interior can be warmer
than $T_c$ even though the normal--superconducting boundaries occur near $T_c$.
Accordingly, $T_{hs}$ is treated as an effective hot-spot temperature and
fitted to the measured $I_r(T_b)$, rather than being fixed at $T_c$.
\cite{dane2022}

First, consider
longitudinal conduction.\cite{tinkham2003} In the symmetric one-dimensional limit, a normal domain of length
$\ell_{hs}=x\ell$, occupying a fraction $x$ of the wire, has resistance $xR_n$, while the
remaining superconducting material forms two arms of length $(1-x)\ell/2$ that conduct to
contacts held at the bath temperature $T_b$. Each arm conducts
$2\kappa A\,(T_{hs}-T_b)/[(1-x)\ell]$ away from the hot spot, so the two together remove
${4\kappa A\,(T_{hs}-T_b)/[(1-x)\ell]}$; equating this to the Joule power $I^2 x R_n$ gives
${x(1-x) = 4\kappa\,A\,(T_{hs}-T_b)/(\ell I^2 R_n)}$. Since $x(1-x)$ is maximized at $x = 1/2$,
where it equals $1/4$, the smallest current that can sustain a stationary hot spot is
\begin{equation}
    \small
  I_r^{\mathrm{cond}} \;=\; 4\sqrt{\frac{\kappa A\,(T_{hs}-T_b)}{\ell R_n}}
       \;=\; \frac{4W}{\ell}\,\sqrt{\frac{\kappa\,t\,(T_{hs}-T_b)}{R_\square}},
  \label{eq:tinkham}
\end{equation}
where $A = Wt$ is the wire cross-sectional area, ${R_n = R_\square \ell/W}$ is the normal-state
resistance of the full wire, $R_\square$ is the sheet resistance, and $\kappa$ is an effective
along-film thermal conductivity over the temperature range of the superconducting arms
(Fig.~\ref{fig:device}). If
$\kappa$ varies with temperature, the factor
$\kappa(T_{hs}-T_b)$ in Eq.~\eqref{eq:tinkham} is replaced by
$\int_{T_b}^{T_{hs}}\kappa(T)\,dT$.\cite{tinkham2003}
Equation~\eqref{eq:tinkham} therefore represents the constant- or
temperature-averaged-$\kappa$ limit of longitudinal cooling, rather than a
universal temperature dependence for heat conduction along the wire.

Second, vertical phonon-mediated cooling removes heat from the film into the surrounding
stack,\cite{sbt1974,dane2022} with an effective areal flux
$\Sigma_K(T^n-T_b^n)$ and ${n=4}$. For a normal segment of length $\ell_{hs}$, both the Joule power
$I^2 R_\square\,\ell_{hs}/W$ and the vertical cooling power $\Sigma_K W \ell_{hs}\,(T_{hs}^4-T_b^4)$ scale
with $\ell_{hs}$; the segment length therefore cancels, giving:
\begin{equation}
  I_r^{\mathrm{int}} \;=\; W\sqrt{\frac{\Sigma_K\,(T_{hs}^4 - T_b^4)}{R_\square}}.
  \label{eq:kapitza}
\end{equation}
Heat taking this route leaves the film through two parallel vertical paths: downward through the amorphous dielectric toward the silicon handle, and upward through the doped polysilicon cap. The downward path is expected to dominate, as the upward one meets the poor sub-kelvin conductivity of the polysilicon and pre-metal dielectric and ends largely in floating metal fill rather than a dedicated heat sink. As used here, \(\Sigma_K\) is therefore an \emph{effective} stack-level coefficient per unit TiN area: it folds several series bottlenecks---multiple interfaces, transport through amorphous dielectric, and the absence of a dedicated upper heat sink---into a single number, and is not the microscopic thermal-boundary conductance of any individual interface.
\begin{figure}[tb]
  \includegraphics[width=\columnwidth]{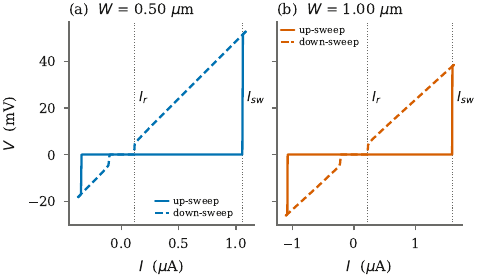}
  \caption{Hysteretic current--voltage characteristics of TiN nanowires,
    $\ell = \SI{50}{\micro\metre}$, at $T_b = \SI{0.15}{\kelvin}$.
    (a) $W = \SI{0.5}{\micro\metre}$: the up-sweep switches at
    $I_{sw} = \SI{1.055}{\micro\ampere}$ and the down-sweep retraps at
    $I_r = \SI{0.109}{\micro\ampere}$.
    (b) $W = \SI{1.0}{\micro\metre}$: $I_{sw} = \SI{1.590}{\micro\ampere}$,
    $I_r = \SI{0.219}{\micro\ampere}$. Vertical dotted lines mark the extracted
    currents. Each $I_{sw}$ above is that of the single trace shown; because individual
    sweeps scatter downward through premature switching, the text instead quotes the
    maximum over the $T_b \le \SI{0.6}{\kelvin}$ plateau.
    Thresholds are taken on the positive branch; $I_r$ agrees between polarities
    to within \SI{4}{\percent}. }
  \label{fig:iv}
\end{figure}


The TiN films are found in poly-resistor structures in a GlobalFoundries
\SI{22}{\nano\metre} FD-SOI process.\cite{swift2025} The wires studied here have $T_c = \SI{1.2}{\kelvin}$ and
an assumed film thickness $t = \SI{3}{\nano\metre}$.\cite{QM_SC_PHASE,DickJames} Four-point $I$--$V$
measurements were performed on two devices of
width $W = 0.5$ and \SI{1.0}{\micro\metre}, both of length $\ell = \SI{50}{\micro\metre}$,
on one die, at bath temperatures ranging from 0.15 to
\SI{1.19}{\kelvin}.  The measured normal-state sheet resistance is
$R_\square = 492~\Omega/\square$.

$I_{sw}$ and $I_r$ are defined as the largest positive currents within the zero-voltage band on the up-sweep and down-sweep branches respectively; the band is set to
\SI{2}{\percent} of the full-scale voltage. $I_r$ is reproducible sweep to sweep. $I_{sw}$ is
not: it varies because some individual sweeps switch prematurely at lower currents. The $I_{sw}$ values
quoted below are therefore the maxima measured over the low-temperature plateau, $T_b \le \SI{0.6}{\kelvin}$.

Figure~\ref{fig:iv} shows that the wires enter a self-sustaining resistive state,
hysteretic between the switching and retrapping currents. The narrow device gives
$I_{sw} = \SI{1.065}{\micro\ampere}$ and
$I_r = \SI{0.109}{\micro\ampere}$, a ratio of 9.8. The wide device gives
$I_{sw} = \SI{1.615}{\micro\ampere}$ and $I_r = \SI{0.219}{\micro\ampere}$, a ratio of 7.4.

\begin{figure}[tb]
  \includegraphics[width=\columnwidth]{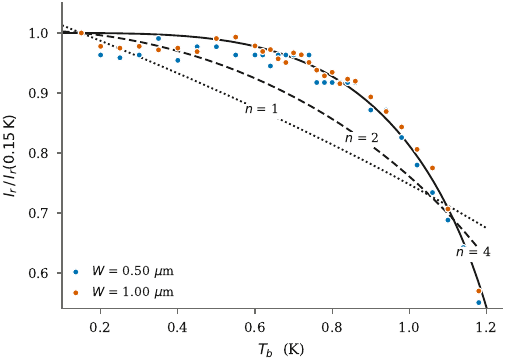}
  \caption{$I_r$ normalized to its \SI{0.15}{\kelvin} value
    against $T_b$, for both $\ell = \SI{50}{\micro\metre}$ devices, with
    $I_r = C\,(T_{hs}^{\,n}-T_b^{\,n})^{1/2}$
    fitted in the prefactor $C$ and the hot-spot temperature $T_{hs}$ and normalized at
    the same point, for $n = 1$ (constant-$\kappa$ longitudinal-conduction limit), 2, and 4 (conventional interfacial phonon-radiation model). Among the fixed exponents shown, $n = 4$ best follows the
    data over the full range.}
  \label{fig:cooling}
\end{figure}

Averaged over the measured bath-temperature range, $T_b=\SI{0.15}{\kelvin}$--\SI{1.19}{\kelvin}, the measured ratio \(I_r(\SI{1.0}{\micro\metre})/I_r(\SI{0.5}{\micro\metre})\) is \(2.036\pm0.034\), consistent with the linear scaling in \(W\) shared by \(I_r^{\mathrm{cond}}\) and \(I_r^{\mathrm{int}}\).
The temperature dependence, by contrast, differs between the two models: both give
$I_r \propto (T_{hs}^{\,n} - T_b^{\,n})^{1/2}$, with $n=1$ for the constant-$\kappa$ longitudinal-conduction limit
and $n=4$ for the conventional interfacial phonon-radiation model. Fitting
$I_r = C\,(T_{hs}^{\,n}-T_b^{\,n})^{1/2}$ with $C$ and $T_{hs}$ both free parameters
decisively rejects the $n=1$ constant-$\kappa$
longitudinal-conduction limit, while $n = 4$
(interfacial phonon radiation) follows the measurements over the full bath-temperature range, $T_b=\SI{0.15}{\kelvin}$--\SI{1.19}{\kelvin}
(Fig.~\ref{fig:cooling}). Allowing the exponent itself to float returns
$n=4.81(24)$ and $4.96(17)$ for the two devices. An explicit $n=5$
fit follows the temperature dependence better than $n=4$
(RMS residual 0.9--1.5\% versus 1.4--1.8\%). Nevertheless, the two
parameterizations give areal cooling fluxes that differ by less than
\SI{3}{\percent} at $T_b=\SI{0.15}{\kelvin}$, indicating that the inferred
magnitude of heat removal is insensitive to this choice. The $n=4$ form
is retained as the conventional phonon-radiation parameterization for
direct comparison with Ref.~\onlinecite{dane2022} and the acoustic- and
diffuse-mismatch benchmarks. Because the TiN film is ultrathin, reduced-dimensional phonon modes could in
principle modify the temperature dependence of the conventional three-dimensional
phonon-radiation model. The fitted exponent $n\approx4.8$--$5.0$ shows no
evidence for a reduced-exponent regime over the measured temperature range,
although the exponent alone does not uniquely identify the microscopic
thermal bottleneck.

\begin{figure}[tb]
  \includegraphics[width=\columnwidth]{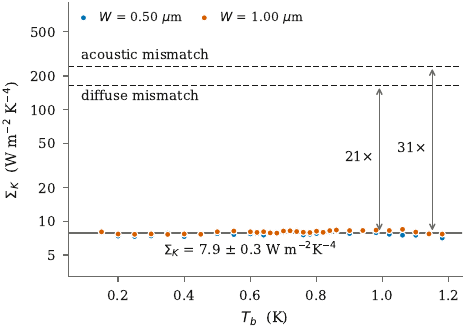}
  \caption{Effective cooling (Kapitza) coefficient $\Sigma_K$ extracted point by point from
    Eq.~\eqref{eq:kapitza} for both $\ell = \SI{50}{\micro\metre}$ devices.
    The 62 extractions below $T_c$
    give one temperature-independent value,
    $\SI{7.9(3)}{\watt\per\metre\squared\per\kelvin\tothe{4}}$ (solid line), $31(21)\times$ below the tabulated acoustic- (diffuse-) mismatch value for a single clean
    TiN/SiO$_2$ interface (dashed lines).\cite{dane2022}}
  \label{fig:sigma}
\end{figure}

Setting $I_r^{\mathrm{int}}$ equal to the measured $I_r$ and rearranging
Eq.~\eqref{eq:kapitza} gives
\[
\Sigma_K =
\frac{I_r^2 R_\square}
{W^2\left(T_{hs}^4-T_b^4\right)}.
\]
Here, $T_{hs}$ is fitted independently for the two devices in the $n=4$
model, yielding $T_{hs}=\SI{1.304}{\kelvin}$ and
$\SI{1.314}{\kelvin}$ for $W=\SI{0.50}{\micro\metre}$ and
$\SI{1.00}{\micro\metre}$, respectively. Their mean,
$T_{hs}=\SI{1.309}{\kelvin}$, is then held fixed when $\Sigma_K$ is
evaluated point-by-point at each measured $T_b$ for both devices. The 62
points below $T_c$ yield
$\SI{7.9(3)}{\watt\per\metre\squared\per\kelvin\tothe{4}}$, with no
resolved systematic dependence on temperature or width
(Fig.~\ref{fig:sigma}). The
parenthetical uncertainty is the sample standard deviation of the
point-by-point values, while the corresponding standard error is
approximately
$\SI{0.04}{\watt\per\metre\squared\per\kelvin\tothe{4}}$.

Equivalently, comparison of the fitted form
\[
I_r=C\left(T_{hs}^4-T_b^4\right)^{1/2},
\]
with Eq.~\eqref{eq:kapitza} gives
\[
C=W\sqrt{\frac{\Sigma_K}{R_\square}},
\qquad
\Sigma_K=\frac{C^2R_\square}{W^2}.
\]
The fitted prefactors independently give
$\Sigma_K=\SI{7.90}{\watt\per\metre\squared\per\kelvin\tothe{4}}$
and
$\SI{7.88}{\watt\per\metre\squared\per\kelvin\tothe{4}}$
for the two devices, consistent with the point-by-point mean. Thus, the
point-by-point extraction also provides a consistency check that the
effective cooling coefficient is independent of $T_b$ and $W$. 

For TiN on SiO$_2$, Ref.~\onlinecite{dane2022} tabulates ideal thermal-boundary coefficients of
\SI{243}{\watt\per\metre\squared\per\kelvin\tothe{4}} for the acoustic-mismatch
model\cite{Kaplan1979-AMM} and \SI{164}{\watt\per\metre\squared\per\kelvin\tothe{4}} for
the diffuse-mismatch model.\cite{swartz1989} These single-interface values exceed the effective cooling
coefficient extracted for the buried CMOS stack by factors of $31(21)\times$
for the acoustic- (diffuse-) mismatch model. A fairer clean-limit
benchmark for a buried film is the dominant downward
stack of Fig.~\ref{fig:device}: combining the
thermal resistances of the layers and interfaces in series gives ideal effective coefficients
of \SI{62.8}{\watt\per\metre\squared\per\kelvin\tothe{4}} (acoustic-mismatch) and
\SI{51.4}{\watt\per\metre\squared\per\kelvin\tothe{4}} (diffuse-mismatch), corresponding to factors of $8.0(6.5)\times$ above the measured $\Sigma_K$ for the acoustic- (diffuse-) mismatch model. Dane \textit{et al.} fitted $\Sigma_K$
for 15 single-layer NbN nanowires on six substrates and obtained values
consistent with the acoustic-mismatch prediction.\cite{dane2022} The buried CMOS TiN film measured
here lies a factor of 12 below the lowest effective cooling coefficient among those 15
directly comparable wires.

What has been established so far applies to long wires. For $\ell = \SI{50}{\micro\metre}$, $I_r$ is
set by vertical phonon-mediated cooling into the surrounding stack (Fig.~\ref{fig:sigma}).
Conduction to the contacts scales with the cross-section $W \times t$ while vertical cooling
scales with the footprint $W \times \ell$, so their relative importance scales primarily with
$t/\ell$, making wire length the experimental control parameter for the crossover. The data that
follow instead vary $\ell$ at fixed $T_b$,
spanning $\ell = 1$--\SI{50}{\micro\metre} at $W = 0.50$ and \SI{1.00}{\micro\metre},
measured at $T_b = \SI{0.15}{\kelvin}$.\cite{swift-thesis}

Where the crossover falls is set by the film thermal conductivity. It is estimated using the expressions of Ref.~\onlinecite{dane2022}, combining a Bardeen--Rickayzen--Tewordt electronic contribution\cite{bardeen1959} with the normal-state Wiedemann--Franz conductivity and a Casimir-limit phonon contribution. Given a measured sheet resistance $R_\square$, the inferred normal-state resistivity is $\rho_n=R_\square t$, which is in turn used for the
Wiedemann--Franz electronic thermal conductivity $\kappa_n=L_0T/\rho_n = L_0T/R_\square t$. For $t=\SI{3}{\nano\metre}$, the resulting
model-based estimate is $\kappa(T_c)=\SI{19.9}{mW/(m.K)}$, of the order measured in other superconducting thin films at sub-Kelvin temperatures.\cite{feshchenko2017}

An independent retrapping-current analysis by Swift, performed assuming $t\sim\SI{5}{\nano\metre}$, yielded
$\kappa\approx\SI{9}{\milli\watt\per\metre\per\kelvin}$.\cite{swift-thesis} Rescaling that estimate to $t=\SI{3}{\nano\metre}$ while keeping the measured $R_\square$ gives $\kappa\approx\SI{15}{\milli\watt\per\metre\per\kelvin}$, of the same order as the present model-based value. Since the electronic contribution dominates, the product $\kappa\,t$ entering Eqs.~\eqref{eq:eta} and \eqref{eq:lstar} is nearly independent of the assumed film thickness.


Treating longitudinal conduction and vertical cooling together requires the introduction of the
linearized thermal healing length\cite{Kerman2007,Vodolazov2017} $\eta$,
which is the distance heat spreads along the film before it escapes vertically into the surrounding stack:
\begin{equation}
  \eta \;=\; \sqrt{\frac{\kappa\,t}{n\,\Sigma_K\,T_{hs}^{\,n-1}}} \;=\; \SI{0.92}{\micro\metre},
  \label{eq:eta}
\end{equation}
with the measured $\Sigma_K$, exponent $n$, and computed $\kappa$. A hot spot much longer than $\eta$
loses its heat vertically into the surrounding stack and never reaches the contacts. One shorter than a few
$\eta$ loses an appreciable fraction longitudinally, out through the contacts at either
end, and needs more current to sustain itself. $I_r(\ell)$ therefore increases at shorter
lengths and plateaus at longer ones. Combining the two limiting loss terms into a single
compact interpolation gives the two-channel form
$I_r^{\mathrm{tot}}$:
\begin{align}
  I_r^{\mathrm{tot}}(\ell) &\;=\; W\sqrt{\frac{\Sigma_K (T_{hs}^4-T_b^4)}{R_\square}}\;
                \sqrt{1+\left(\frac{\ell^*}{\ell}\right)^{2}},
  \label{eq:length}\\[2pt]
  \ell^*    &\;=\; 2\sqrt{\frac{\kappa\,t\,(T_{hs}-T_b)}{\Sigma_K (T_{hs}^4-T_b^4)}},
  \label{eq:lstar}
\end{align}
with $\ell^* = 3.8\,\eta = \SI{3.5}{\micro\metre}$ estimated from the $\Sigma_K$ and $\kappa$
fixed above. Equation~\eqref{eq:length} is phenomenological: it lumps the hot spot into a
single temperature $T_{hs}$ rather than solving for the profile along the wire, and the
prefactor in Eq.~\eqref{eq:lstar} is a heuristic $O(1)$ constant carried over from the
single-channel hot-spot profile, so $\ell^*$ is an order-of-magnitude prediction.
For $\ell \gg \ell^*$ it returns the length-independent $I_r^{\mathrm{int}}$ of Eq.~\eqref{eq:kapitza}. For $\ell \ll \ell^*$ the contacts dominate and it recovers the
$\ell^{-1}$ scaling of Eq.~\eqref{eq:tinkham}, with an amplitude that depends on the hot-spot profile assumed.

\begin{figure}[t]
  \centering
  \includegraphics[width=\columnwidth]{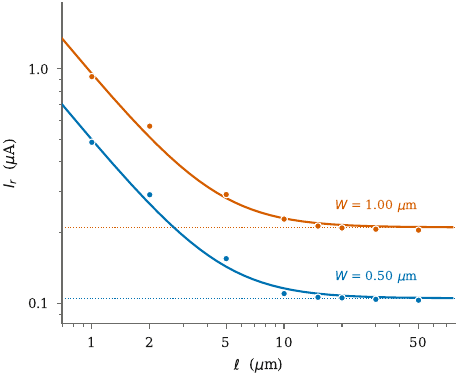}
  \caption{Length dependence of $I_r$ across the TiN test structures of
    Ref.~\onlinecite{swift-thesis}, measured at $T_b = \SI{0.15}{\kelvin}$. Markers are measured data.
    Solid curves are Eq.~\eqref{eq:length} with amplitude and $\ell^*$ fitted per width;
    dotted lines are the corresponding $\ell \gg \ell^*$ interfacial floors.}
  \label{fig:length}
\end{figure}

Equation~\eqref{eq:length} is applied to these structures in Fig.~\ref{fig:length}.
The data fall steeply over the first decade of length, roll over
at a few micrometres, and are flat by \SI{30}{\micro\metre};
the two widths run parallel, separated by their width ratio.
Equation~\eqref{eq:length} is fitted to each width with
two free parameters, the plateau current $I_\infty$ and the
crossover length $\ell^*$. The fits return
$I_\infty = 0.105$ and \SI{0.211}{\micro\ampere} and $\ell^* = 4.6$ and
\SI{4.4}{\micro\metre}. The plateau currents agree with the $I_\infty = 0.108$ and
\SI{0.217}{\micro\ampere} obtained from the \SI{50}{\micro\metre} measurements, and the
fitted $\ell^*$ agrees with the predicted $\ell^* \approx \SI{3.5}{\micro\metre}$ to within the
$O(1)$ prefactor of Eq.~\eqref{eq:lstar}.
The unusually long crossover follows directly from the weak vertical cooling: at
fixed $\kappa t$, $T_{hs}$, and $T_b$, Eq.~\eqref{eq:lstar} gives
$\ell^* \propto \Sigma_K^{-1/2}$. Three cooling coefficients therefore set three
crossover lengths. The measured
$\Sigma_K=\SI{7.9}{\watt\per\metre\squared\per\kelvin\tothe{4}}$ gives
$\ell^*=\SI{3.5}{\micro\metre}$. The ideal buried stack,
$62.8$($51.4$)\,\si{\watt\per\metre\squared\per\kelvin\tothe{4}}, gives
$1.2$($1.4$)\,\si{\micro\metre}. A single clean interface,
$243$($164$)\,\si{\watt\per\metre\squared\per\kelvin\tothe{4}}, gives
$0.6$($0.8$)\,\si{\micro\metre}, quoted throughout for the acoustic-
(diffuse-) mismatch model.
This benchmark is derived here from Eq.~\eqref{eq:lstar}, not a crossover length measured by
Ref.~\onlinecite{dane2022}. An otherwise similar film with conventional strong vertical thermal
coupling would therefore reach its $I_r(\ell)$ rollover at a wire length about six times shorter: stronger vertical
cooling removes heat within a shorter healing length $\eta$, which shortens $\ell^*$ and moves the
knee in $I_r(\ell)$ to shorter wires. The 2-D substrate model gives a characteristic lateral spreading length
of approximately \SI{234}{\micro\metre}, well above the
\SIrange{1}{50}{\micro\metre} wire-length range studied here. The
substrate therefore acts as an approximately uniform thermal bath over the measured devices, and the few-micrometer crossover is instead set by the competition between longitudinal conduction along the TiN film
and vertical cooling into the surrounding stack.


The extracted $\Sigma_K$ depends on the modeling conventions used to reduce the
data, chiefly the hot-spot-temperature assignment. The extraction relation
constrains only the product
$\Sigma_K(T_{hs}^4-T_b^4)$, so the inferred coefficient scales inversely with
the assumed thermal driving term. The headline value
$\Sigma_K=\SI{7.9(3)}{\watt\per\metre\squared\per\kelvin\tothe{4}}$
follows the fitted-$T_{hs}$ convention of Ref.~\onlinecite{dane2022}, for which
the $n=4$ fit of Fig.~\ref{fig:cooling} gives
$T_{hs}\approx\SI{1.31}{\kelvin}$, slightly above
$T_c=\SI{1.2}{\kelvin}$. Fixing $T_{hs}=T_c$ and refitting the two devices
instead gives $\Sigma_K\approx\SI{12}{\watt\per\metre\squared\per\kelvin\tothe{4}}$.

Wire width constitutes a separate systematic uncertainty because
$\Sigma_K\propto W^{-2}$; for example, a \SI{10}{\percent} reduction in the
assumed width would increase the inferred coefficient by approximately
\SI{23}{\percent}. The reported extraction nevertheless uses the drawn device
widths throughout and does not assume a narrower effective current-carrying
channel. Thus, the same retrapping data yield
$\Sigma_K=7.9$ or approximately
$12$~\si{\watt\per\metre\squared\per\kelvin\tothe{4}},
depending on the hot-spot-temperature convention, reflecting modeling
uncertainty rather than distinct physical cooling coefficients. The latter
value remains approximately $20(14)\times$ below the acoustic-
(diffuse-) mismatch value for a single clean TiN/SiO$_2$ interface, compared
with $31(21)\times$ for the headline fitted-$T_{hs}$ value. The inferred
coefficients also remain below the range reported for NbN by
Ref.~\onlinecite{dane2022}. Electron--phonon nonequilibrium was also assessed using the
two-temperature model of Ref.~\onlinecite{dane2022}, in which the
electron--phonon heat flux is
$t\Sigma_{e\text{-}ph}(T_e^5-T_{ph}^5)$, with
$\Sigma_{e\text{-}ph}$ parameterized following
Refs.~\onlinecite{kaplan1976,dane2022}.
Using an order-of-magnitude TiN coupling estimate together with the
measured $\Sigma_K$, $t=\SI{3}{\nano\metre}$, and the fitted
$T_{hs}\approx\SI{1.309}{\kelvin}$ gives
$T_e-T_{ph}\approx\SI{31}{\milli\kelvin}$ and a differential
electron--phonon conductance approximately ten times the film-to-bath
conductance. These numerical values are estimates for the present film. Published TiN
electron--phonon relaxation times vary substantially,\cite{kardakova2013}
so this calculation is used only as an order-of-magnitude consistency
check.

In summary, TiN nanowires buried in the FEOL beneath the BEOL stack of a commercial CMOS process
sustain self-heated hot spots with hysteresis ratios of 7.4--9.8. Transient hot-spot
formation underpins nanowire detection; controlled hysteresis, nanowire logic and memory.
Retrapping in the long wires
is governed by vertical phonon-mediated cooling into the surrounding stack rather than by
longitudinal conduction to the contacts. A single effective cooling coefficient,
$\Sigma_K = \SI{7.9(3)}{\watt\per\metre\squared\per\kelvin\tothe{4}}$,
describes the data under the conventional $n = 4$ parameterization.
Over a fiftyfold range of length, $I_r$ follows neither
the $\ell^{-1}$ longitudinal-conduction limit nor the $\ell^{0}$ vertical-cooling limit, but crosses between them
at a few micrometres (fitted \SIrange{4.4}{4.6}{\micro\metre}), near the two-channel crossover $\ell^* \approx \SI{3.5}{\micro\metre}$
predicted from the measured $\Sigma_K$ and the computed $\kappa$. The unusually small effective $\Sigma_K$ therefore
explains both the low long-wire retrapping current and the extended length scale over which
contact cooling remains important.

\begin{acknowledgments}
This work was supported by the U.S. Department of Energy, Office of Science, Advanced Scientific Computing Research (ASCR), High Energy Physics (HEP) and Basic Energy Sciences (BES), under Contract No. DE-AC02-06CH11357 at Argonne. Work performed at the Advanced Photon Source, a U.S. Department of Energy Office of Science User Facilities (SUF). Work at University of Chicago is supported by the Divisions of Chemistry (CHE) and Materials Research (DMR), National Science Foundation, under grant numbers NSF/CHE-1834750 and NSF/CHE-2335833. M.F.G.Z. acknowledges a UKRI Future Leaders Fellowship [MR/V023284/1]. A.G.-S. acknowledges an Industrial Fellowship from the Royal Commission for the Exhibition of 1851. This work was primarily supported by the CMOS+X project funded by DOE's Genesis Mission. 
\end{acknowledgments}

\section*{Author Declarations}

\subsection*{Conflict of Interest}
The authors have no conflicts to disclose.

\subsection*{Author Contributions}

\noindent
\textbf{Mohamed Gharib:}
Methodology (equal); Software (equal); Formal analysis (lead);
Writing -- original draft (equal); Writing -- review \& editing (equal);
Visualization (lead).
\textbf{Leonid Popryho:}
Software (equal); Validation (lead); Writing -- review \& editing (equal).
\textbf{Salma Abdelzaher:}
Writing -- review \& editing (equal).
\textbf{Tejas Guruswamy:}
Methodology (equal); Writing -- review \& editing (equal);
Supervision (equal); Funding acquisition (equal).
\textbf{Tomas Polakovic:}
Methodology (equal); Writing -- review \& editing (equal);
Funding acquisition (equal).
\textbf{Umeshkumar Patel:}
Methodology (equal); Writing -- review \& editing (equal);
Funding acquisition (equal).
\textbf{Orlando Quaranta:}
Methodology (equal); Writing -- review \& editing (equal);
Funding acquisition (equal).
\textbf{Thomas Cecil:}
Methodology (equal); Writing -- review \& editing (equal);
Funding acquisition (equal).
\textbf{Clarence Chang:}
Methodology (equal); Writing -- review \& editing (equal);
Funding acquisition (equal).
\textbf{Yu-Sheng Chen:}
Investigation (equal); Writing -- review \& editing (equal).
\textbf{Thomas H. Swift:}
Formal analysis (lead); Investigation (equal); Data curation (lead);
Writing -- review \& editing (equal).
\textbf{Grayson M. Noah:}
Investigation (equal); Writing -- review \& editing (equal).
\textbf{Alberto Gomez-Saiz:}
Investigation (equal); Writing -- review \& editing (equal);
Visualization (equal).
\textbf{John J. L. Morton:}
Investigation (equal); Writing -- review \& editing (equal);
Visualization (equal).
\textbf{M. Fernando Gonzalez-Zalba:}
Investigation (equal); Writing -- review \& editing (equal);
Visualization (equal).
\textbf{Antonino Miceli:}
Conceptualization (equal); Writing -- original draft (equal);
Writing -- review \& editing (equal); Supervision (equal);
Project administration (equal); Funding acquisition (equal).
\textbf{Inna Partin-Vaisband:}
Conceptualization (equal); Writing -- review \& editing (equal);
Supervision (equal); Project administration (equal);
Funding acquisition (equal).

\section*{Data Availability}
The data that support the findings of this study are available from the corresponding
authors upon reasonable request.

\bibliography{refs}

@article{Goltsman,
    author = {Gol’tsman, G. N. and Okunev, O. and Chulkova, G. and Lipatov, A. and Semenov, A. and Smirnov, K. and Voronov, B. and Dzardanov, A. and Williams, C. and Sobolewski, Roman},
    title = {Picosecond superconducting single-photon optical detector},
    journal = {Applied Physics Letters},
    volume = {79},
    number = {6},
    pages = {705-707},
    year = {2001},
    month = {08},
    issn = {0003-6951},
    doi = {10.1063/1.1388868},
    url = {https://doi.org/10.1063/1.1388868},
}

@article{Lita:08,
author = {Adriana E. Lita and Aaron J. Miller and Sae Woo Nam},
journal = {Opt. Express},
number = {5},
pages = {3032--3040},
publisher = {Optica Publishing Group},
title = {Counting near-infrared single-photons with 95\% efficiency},
volume = {16},
month = {Mar},
year = {2008},
url = {https://opg.optica.org/oe/abstract.cfm?URI=oe-16-5-3032},
doi = {10.1364/OE.16.003032},
}

@article{Vodolazov2017,
  title = {Single-Photon Detection by a Dirty Current-Carrying Superconducting Strip Based on the Kinetic-Equation Approach},
  author = {Vodolazov, D. Yu.},
  journal = {Phys. Rev. Appl.},
  volume = {7},
  issue = {3},
  pages = {034014},
  numpages = {19},
  year = {2017},
  month = {Mar},
  publisher = {American Physical Society},
  doi = {10.1103/PhysRevApplied.7.034014},
  url = {https://link.aps.org/doi/10.1103/PhysRevApplied.7.034014}
}

@article{Kerman2007,
    author = {Kerman, Andrew J. and Dauler, Eric A. and Yang, Joel K. W. and Rosfjord, Kristine M. and Anant, Vikas and Berggren, Karl K. and Gol’tsman, Gregory N. and Voronov, Boris M.},
    title = {Constriction-limited detection efficiency of superconducting nanowire single-photon detectors},
    journal = {Applied Physics Letters},
    volume = {90},
    number = {10},
    pages = {101110},
    year = {2007},
    month = {03},
    issn = {0003-6951},
    doi = {10.1063/1.2696926},
    url = {https://doi.org/10.1063/1.2696926},
}

@article{Kaplan1979-AMM,
  author  = {Kaplan, S. B.},
  title   = {Acoustic matching of superconducting films to substrates},
  journal = {Journal of Low Temperature Physics},
  year    = {1979},
  volume  = {37},
  pages   = {343--365},
  doi     = {10.1007/BF00119193}
}

@article{htron,
  title = {Multilayered Heater Nanocryotron: A Superconducting-Nanowire-Based Thermal Switch},
  author = {Baghdadi, Reza and Allmaras, Jason P. and Butters, Brenden A. and Dane, Andrew E. and Iqbal, Saleem and McCaughan, Adam N. and Toomey, Emily A. and Zhao, Qing-Yuan and Kozorezov, Alexander G. and Berggren, Karl K.},
  journal = {Phys. Rev. Appl.},
  volume = {14},
  issue = {5},
  pages = {054011},
  numpages = {12},
  year = {2020},
  month = {Nov},
  publisher = {American Physical Society},
  doi = {10.1103/PhysRevApplied.14.054011},
  url = {https://link.aps.org/doi/10.1103/PhysRevApplied.14.054011}
}

@ARTICLE{orlando-ntron,
  author={Quaranta, Orlando and Marchetti, Stefania and Martucciello, Nadia and Pagano, Sergio and Ejrnaes, Mikkel and Cristiano, Roberto and Nappi, Ciro},
  journal={IEEE Transactions on Applied Superconductivity}, 
  title={Superconductive Three-Terminal Amplifier/Discriminator}, 
  year={2009},
  volume={19},
  number={3},
  pages={367-370},
  url={https://doi.org.10.1109/TASC.2009.2017952}
  }

@article{Berggren-ntron,
author = {McCaughan, Adam N. and Berggren, Karl K.},
title = {A Superconducting-Nanowire Three-Terminal Electrothermal Device},
journal = {Nano Letters},
volume = {14},
number = {10},
pages = {5748-5753},
year = {2014},
URL = {https://doi.org/10.1021/nl502629x}
}

@ARTICLE{TES-APS-ICs,
  author={Guruswamy, Tejas and Gades, Lisa and Miceli, Antonino and Patel, Umeshkumar and Quaranta, Orlando},
  journal={IEEE Transactions on Applied Superconductivity}, 
  title={Beamline Spectroscopy of Integrated Circuits With Hard {X}-Ray Transition Edge Sensors at the {Advanced Photon Source}}, 
  year={2021},
  volume={31},
  number={5},
  pages={1-5},
  doi={10.1109/TASC.2021.3067246}}

@article{swift2025,
  author  = {Swift, Thomas H. and Olivieri, Fabio and Aizpurua-Iraola, Gorka and
             Kirkman, James and Noah, Grayson M. and de Kruijf, Mathieu and
             von Horstig, Felix-Ekkehard and Gomez-Saiz, Alberto and
             Morton, John J. L. and Gonzalez-Zalba, M. Fernando},
  title   = {A superinductor in a deep sub-micron integrated circuit},
  journal = {arXiv:2507.13202},
  year    = {2025},
  doi     = {10.48550/arXiv.2507.13202}
}

@article{natarajan2012,
  author  = {Natarajan, Chandra M. and Tanner, Michael G. and Hadfield, Robert H.},
  title   = {Superconducting nanowire single-photon detectors:
             physics and applications},
  journal = {Supercond. Sci. Technol.},
  volume  = {25},
  number  = {6},
  pages   = {063001},
  year    = {2012},
  doi     = {10.1088/0953-2048/25/6/063001}
}

@article{wittmer1980,
  author  = {Wittmer, M.},
  title   = {{TiN} and {TaN} as diffusion barriers in metallizations to silicon semiconductor devices},
  journal = {Applied Physics Letters},
  volume  = {36},
  number  = {6},
  pages   = {456--458},
  year    = {1980},
  doi     = {10.1063/1.91505}
}

@article{lavoie2003,
  author  = {Lavoie, C. and d'Heurle, F. M. and Detavernier, C. and Cabral, Jr., C.},
  title   = {Towards implementation of a nickel silicide process for {CMOS} technologies},
  journal = {Microelectronic Engineering},
  volume  = {70},
  number  = {2--4},
  pages   = {144--157},
  year    = {2003},
  doi     = {10.1016/S0167-9317(03)00380-0}
}

@article{vissers2010,
  author  = {Vissers, M. R. and Gao, J. and Wisbey, D. S. and Hite, D. A. and Tsuei, C. C. and Corcoles, A. D. and Steffen, M. and Pappas, D. P.},
  title   = {Low loss superconducting titanium nitride coplanar waveguide resonators},
  journal = {Applied Physics Letters},
  volume  = {97},
  number  = {23},
  pages   = {232509},
  year    = {2010},
  doi     = {10.1063/1.3517252}
}

@article{chiu2021,
  author  = {Chiu, Shao-Pin and Tsuei, C. C. and Yeh, Sheng-Shiuan and Zhang, Fu-Chun and Kirchner, Stefan and Lin, Juhn-Jong},
  title   = {Observation of triplet superconductivity in {CoSi$_2$/TiSi$_2$} heterostructures},
  journal = {Sci. Adv.},
  volume  = {7},
  number  = {29},
  pages   = {eabg6569},
  year    = {2021},
  doi     = {10.1126/sciadv.abg6569}
}

@article{tinkham2003,
  author  = {Tinkham, M. and Free, J. U. and Lau, C. N. and Markovi\'{c}, N.},
  title   = {Hysteretic {$I$--$V$} curves of superconducting nanowires},
  journal = {Phys. Rev. B},
  volume  = {68},
  number  = {13},
  pages   = {134515},
  year    = {2003},
  doi     = {10.1103/PhysRevB.68.134515}
}

@article{sbt1974,
  author  = {Skocpol, W. J. and Beasley, M. R. and Tinkham, M.},
  title   = {Self-heating hotspots in superconducting thin-film microbridges},
  journal = {J. Appl. Phys.},
  volume  = {45},
  number  = {9},
  pages   = {4054--4066},
  year    = {1974},
  doi     = {10.1063/1.1663912}
}

@article{dane2022,
  author  = {Dane, Andrew and Allmaras, Jason and Zhu, Di and Onen, Murat and
             Colangelo, Marco and Baghdadi, Reza and Ba{\v{s}}kent, Nergiz and
             Berggren, Karl K.},
  title   = {Self-heating hotspots in superconducting nanowires cooled by
             phonon black-body radiation},
  journal = {Nat. Commun.},
  volume  = {13},
  pages   = {5429},
  year    = {2022},
  doi     = {10.1038/s41467-022-32719-w}
}

@article{swartz1989,
  author  = {Swartz, E. T. and Pohl, R. O.},
  title   = {Thermal boundary resistance},
  journal = {Rev. Mod. Phys.},
  volume  = {61},
  number  = {3},
  pages   = {605--668},
  year    = {1989},
  doi     = {10.1103/RevModPhys.61.605}
}

@article{feshchenko2017,
  author  = {Feshchenko, A. V. and Saira, O.-P. and Peltonen, J. T. and Pekola, J. P.},
  title   = {Thermal conductance of {Nb} thin films at sub-kelvin temperatures},
  journal = {Sci. Rep.},
  volume  = {7},
  pages   = {41728},
  year    = {2017},
  doi     = {10.1038/srep41728}
}

@article{kerman2009,
  author  = {Kerman, Andrew J. and Yang, Joel K. W. and Molnar, Richard J. and
             Dauler, Eric A. and Berggren, Karl K.},
  title   = {Electrothermal feedback in superconducting nanowire
             single-photon detectors},
  journal = {Phys. Rev. B},
  volume  = {79},
  number  = {10},
  pages   = {100509},
  year    = {2009},
  doi     = {10.1103/PhysRevB.79.100509}
}

@article{annunziata2010,
  author  = {Annunziata, A. J. and Quaranta, O and Santavicca, D F. and Casaburi, A. and Frunzio, L and Ejrnaes, M. and Rooks, M. J. and Cristiano, R. and Pagano, S. and Frydman, A. and Prober, D. E.},
  title   = {Reset dynamics and latching in niobium superconducting nanowire single-photon detectors},
  journal = {J. Appl. Phys.},
  volume  = {108},
  number  = {8},
  pages   = {084507},
  year    = {2010},
  doi     = {10.1063/1.3498809}
}

@article{bardeen1959,
  author  = {Bardeen, J. and Rickayzen, G. and Tewordt, L.},
  title   = {Theory of the thermal conductivity of superconductors},
  journal = {Phys. Rev.},
  volume  = {113},
  number  = {4},
  pages   = {982--994},
  year    = {1959},
  doi     = {10.1103/PhysRev.113.982}
}

@article{kaplan1976,
  author  = {Kaplan, S. B. and Chi, C. C. and Langenberg, D. N. and Chang, J. J. and
             Jafarey, S. and Scalapino, D. J.},
  title   = {Quasiparticle and phonon lifetimes in superconductors},
  journal = {Phys. Rev. B},
  volume  = {14},
  number  = {11},
  pages   = {4854--4873},
  year    = {1976},
  doi     = {10.1103/PhysRevB.14.4854}
}

@phdthesis{swift-thesis,
  author  = {Swift, Thomas Hugh},
  title   = {Characterisation of superconductors and quantum dots in
             22nm-node {CMOS} devices},
  school  = {University College London},
  year    = {2025},
  month   = mar,
}

@misc{QM_SC_PHASE,
      title={A Superconducting Phase Transition Single-Electron Transistor}, 
      author={Gorka Aizpurua-Iraola and Thomas H. Swift and Felix-Ekkehard von Horstig and Domenic Prete and James Kirkman and Grayson M. Noah and Fabio Olivieri and Alberto Gomez-Saiz and M. Fernando Gonzalez-Zalba},
      year={2026},
      eprint={2608.27045},
      archivePrefix={arXiv},
      primaryClass={quant-ph},
      url={https://arxiv.org/abs/2608.27045}, 
}

@article{kardakova2013,
    author = {Kardakova, A. and Finkel, M. and Morozov, D. and Kovalyuk, V. and An, P. and Dunscombe, C. and Tarkhov, M. and Mauskopf, P. and Klapwijk, T. M. and Goltsman, G.},
    title = {The electron-phonon relaxation time in thin superconducting titanium nitride films},
    journal = {Applied Physics Letters},
    volume = {103},
    number = {25},
    pages = {252602},
    year = {2013},
    month = {12},
    issn = {0003-6951},
    doi = {10.1063/1.4851235},
    url = {https://doi.org/10.1063/1.4851235},
}

@INPROCEEDINGS{DickJames,
  author={James, Dick},
  booktitle={25th Annual SEMI Advanced Semiconductor Manufacturing Conference (ASMC 2014)}, 
  title={High-k/metal gates in the 2010s}, 
  year={2014},
  volume={},
  number={},
  pages={431-438},
  doi={10.1109/ASMC.2014.6846970}}

@article{Groll_NICK,
    author = {Groll, Nickolas R. and Klug, Jeffrey A. and Cao, Chaoyue and Altin, Serdar and Claus, Helmut and Becker, Nicholas G. and Zasadzinski, John F. and Pellin, Michael J. and Proslier, Thomas},
    title = {Tunneling spectroscopy of superconducting MoN and NbTiN grown by atomic layer deposition},
    journal = {Applied Physics Letters},
    volume = {104},
    number = {9},
    pages = {092602},
    year = {2014},
    month = {03},
    issn = {0003-6951},
    doi = {10.1063/1.4867880},
    url = {https://doi.org/10.1063/1.4867880},
}

@article{CHIU2024348,
title = {Electronic and superconducting properties of CoSi2 films on silicon — An unconventional superconductor with technological potential},
journal = {Chinese Journal of Physics},
volume = {90},
pages = {348-363},
year = {2024},
issn = {0577-9073},
doi = {https://doi.org/10.1016/j.cjph.2024.04.018},
url = {https://www.sciencedirect.com/science/article/pii/S0577907324001540},
author = {Shao-Pin Chiu and Chang-Jan Wang and Yi-Chun Lin and Shun-Tast Tu and Shouray Kumar Sahu and Ruey-Tay Wang and Chih-Yuan Wu and Sheng-Shiuan Yeh and Stefan Kirchner and Juhn-Jong Lin}
}

\end{document}